\documentclass[aip,reprint]{revtex4-1}

\usepackage[utf8]{inputenc}
\usepackage[T1]{fontenc}
\usepackage{rotating}
\usepackage{dcolumn} 
\usepackage{array}
\usepackage{etoolbox}
\usepackage{graphicx}
\usepackage{setspace}
\usepackage{amsmath,amssymb,epstopdf}
\usepackage{subfig}
\usepackage{mathtools}
\usepackage{xcolor}
\usepackage{soul}
\usepackage{enumitem}

\usepackage{float}
\usepackage{hyperref}

\draft 

\begin{document}

\title[]{Empirically extending 1D Child-Langmuir theory to a finite temperature beam}



\author{Jesse M. Snelling}
 \altaffiliation{jesse.snelling@colorado.edu}
 \affiliation{Center for Integrated Plasma Studies, University of Colorado, Boulder, Colorado 80309, USA}

\author{Gregory R. Werner}
 \affiliation{Center for Integrated Plasma Studies, University of Colorado, Boulder, Colorado 80309, USA}

\author{John R. Cary}
 \affiliation{Center for Integrated Plasma Studies, University of Colorado, Boulder, Colorado 80309, USA}
 \affiliation{Tech-X Corporation, 5621 Arapahoe Avenue Suite A, Boulder, Colorado 80303, USA}


\date{\today}

\begin{abstract}
Numerical solutions to the 1D steady-state Vlasov-Poisson system are used to develop a straightforward empirical formula for the electric current density transmitted through a vacuum diode (voltage gap) as a function of gap distance, gap voltage, the injected current density, and the average velocity and temperature of injected particles, as well as their charge and mass.  
This formula generalizes the 1D cold beam Child-Langmuir law (which predicts the maximum transmitted current for mono-energetic particles in a planar diode as a function of gap voltage and distance) to the case where particles are injected with a finite velocity spread. 
Though this case is of practical importance, no analytical solution is known.  
Found by a best-fit to results from particle-in-cell (PIC) simulations, the empirical formula characterizes the current transmitted across the diode for an injected velocity distribution of a drifting Maxwellian. 
It is not meant to yield a precise answer, but approximately characterizes the effect of space charge on transmitted current density over a large input space.  
The formula allows quick quantitative estimation of the effect of space charge in diode-like devices, such as gate-anode gaps in nanoscale vacuum channel transistors.

\end{abstract}


\maketitle 

\section{Introduction}
\label{sec:introduction}

The space charge limited (SCL) current between two electrodes is a problem of great interest with applications to nanoscale vacuum channel transistors,\cite{Bhattacharya2021} semiconductor diodes,\cite{Zhang2017,Yin2020} ion beams,\cite{Chauvin2012} x-ray detectors in fusion devices,\cite{Trosseille2022} magnetron sputtering,\cite{Shon1998} and ion thrusters. \cite{Grondein2016}
The simplest version of the SCL current problem is the 1D planar geometry where charged particles are injected with zero velocity at one electrode and are subsequently accelerated by the applied gap voltage to the other electrode. 
There exists a maximum steady-state current (the SCL current) beyond which the self-fields of the particles in the gap prevent injected particles from entering the gap, precluding a steady-state solution. 
The SCL current for the zero injection velocity case is given by the well known Child-Langmuir (CL) law:

\begin{equation}
\label{eqn:CLlaw}
    J_{CL} =  \frac{2m\epsilon_0}{9q} \left(\frac{2|q\Phi_D|}{m}\right)^{\frac{3}{2}} D^{-2}
\end{equation}
where $J_{CL}$ is the SCL current density, $\epsilon_0$ is the permittivity of free space, $q$ is the charge of the injected particles, $m$ is the mass of the injected particles, $\Phi_D$ is the gap potential, and $D$ is the gap distance.
(We note that Eq. \ref{eqn:CLlaw} only gives the CL law for $q\Phi_D < 0$, that is, an accelerating gap potential, but we use the nonstandard notation including absolute values for later generalization when $q\Phi_D \geq 0$ for the case of finite injection velocity.)
For an injected current density magnitude ($|J_i|$) less than $|J_{CL}|$, the magnitude of the electric field at the injection electrode is reduced by space charge, but particles are still accelerated into the gap.
For $J_i = J_{CL}$, the electric field at the injection electrode goes to zero and no particles enter the gap. 

This problem was first examined by Child\cite{Child1911} and Langmuir.\cite{Langmuir1913} 
Since then, it has been studied extensively to account for associated instabilities,\cite{Birdsall1961, Bridges1963} higher dimensional geometries,\cite{Watrous2001, Luginsland2002} relativistic regimes,\cite{Greenwood2016} quantum regimes,\cite{Ang2006} and time-dependent boundary conditions.\cite{Griswold2010}
There exists a simple analytical extension\cite{Akimov2001} for the SCL current density for the case of particles injected with the fixed velocity $v_i>0$ (a cold beam):

\begin{align}
\label{eqn:SCLgeneralizedUnsimplified}
    J_{cb} &= \frac{2m\epsilon_0}{9q} \left(v_i + \sqrt{v_i^2 - \frac{2q\Phi_D}{m}}\right)^3  D^{-2} \\ 
    \label{eqn:SCLgeneralized}
    &=  J_{CL}  \left[ \beta + \sqrt{ \beta^2 - \frac{q\Phi_D}{|q\Phi_D|}} \right]^3
\end{align}
where $J_{CL}$ is given in Eq. \ref{eqn:CLlaw} and $\beta \equiv \left(mv_i^2/ 2|q\Phi_D|\right)^{\frac{1}{2}}$.
We note that $J_{cb}$ is monotonic in $\beta$. 
Additionally, if $v_i \to 0$ ($\beta \to 0$) then $J_{cb} \to J_{CL}$; thus $\beta$ quantifies the effect of injection velocity on the SCL current.
When $mv_i^2/2 < q\Phi_D$, particles are energetically forbidden from crossing the gap and we see that Eq. \ref{eqn:SCLgeneralized} yields a complex value.

A natural next step in extending the 1D diode SCL current problem is to include a thermal velocity spread, $v_{\rm th}$, in the description of the injected velocity distribution, which allows the case where only a portion of the particles are reflected by space charge.
However, a general analytical solution to this problem has not yet been presented. 
In this paper, we utilize simulation to provide an approximate formula for the transmitted current density as a function of  particle and gap characteristics, taking into account the effect a thermal particle distribution has on space charge.

We will begin with a precise definition of the finite temperature (warm) beam 1D diode problem in Section \ref{sec:definition}. 
In Section \ref{sec:Reduction}, we reduce the large parameter space to just three dimensions, which makes a numerical scan over the input space feasible. 
In Section \ref{sec:PICsimulation}, we describe the individual simulations. 
The procedure for numerically scanning over the parameter space and fitting the simulation results is presented in Section \ref{sec:fitting}. 
The model parameters along with the model's performance are presented in Section \ref{sec:results}. 
Finally, we provide a brief summary of the paper in Section \ref{sec:summary}. 
The appendix gives a recipe for straightforward implementation of the empirical model in general and in the simplified special case of a ``nearly'' cold beam.

\section{Problem Definition}
\label{sec:definition}

The aim of this paper is to find steady-state solutions of the one-species 1D Vlasov-Poisson system for phase space density $f(x,v,t)$ and electric potential $\Phi(x,t)$ with the distribution of injected current given by $j_i(v)$ in a gap of length $D$ with applied voltage $\Phi_D$. The equations describing such a system are:
\begin{subequations}
\label{eqn:vlasovpoisson}
\begin{alignat}{4}
  \text{(Vlasov)} && \partial_t f &= && -v \partial_x f + \frac{q}{m}(\partial_x \Phi )(\partial_v f),  & \\
  \text{(Poisson)} && -\partial_x^2 \Phi &= && \frac{q}{\epsilon_0} \int f dv,  & \\
  \text{(Left BC)} && \Phi(0,t) &= && 0, & \\
  \text{(Right BC)} && \Phi(D,t) &= && \Phi_D, & \\
  && j(0,v>0,t) &= && q v f(x=0,v>0,t) = j_i(v), &
\end{alignat}
\end{subequations}
where the final equation gives the current density boundary condition (BC). 
We impose particle absorption boundaries at the electrodes at $x=0$ or $x=D$.
Nominally, this problem has inputs $q$, $m$, $D$, $\Phi_D$, and $j_i(v)$ with the output $J_t$, the steady-state current density that is transmitted to $x=D$.

While the Vlasov-Poisson system above describes the problem for arbitrary $j_i(v)$, for the sake of practicality, we restrict our study to a single distribution function. 
A Maxwellian distribution is common in many applications and is a standard choice to capture the first and second moments of a distribution when modeling.
In this paper, we consider only the constant injected current density distribution function $j_i(v)$ of the form of a truncated drifting Maxwellian with drift velocity $v_D>0$, thermal velocity $v_{\rm th}>0$, and injection current density $J_i$:

\begin{equation}
\label{eqn:maxwellian}
    j_i(v|J_i,v_D,v_{\rm th}) =  \frac{J_i \, v}{v_{\rm norm} v_{\rm th} \sqrt{  2 \pi }} \exp{\left( \frac{-\left( v-v_D \right)^2}{2 v_{\rm th}^2} \right)},
\end{equation}
where $v_{\rm norm}$ is given by:
\begin{align}
\label{eqn:vnorm}
    v_{\rm norm}(v_D,v_{\rm th}) &=  \int_{0}^\infty \frac{v}{v_{\rm th} \sqrt{  2 \pi }}\exp{\left( \frac{-\left( v-v_D \right)^2}{2 v_{\rm th}^2} \right)} \, dv \notag\\
    &= \frac{v_D}{2} \rm{erfc}\left(-\frac{v_D}{\sqrt{2}v_{\rm th}}\right) + \frac{v_{\rm th}}{\sqrt{2\pi}}\exp{\left(-\frac{v_D^2}{2v_{\rm th}^2}\right)}.
\end{align}
Because the Maxwellian is truncated, the mean injection velocity $v_i$ differs from $v_D$. 
In principle, $v_D$ can be zero or negative, though we do not explore these cases. 
Regardless, we find it more convenient to work with $v_i$ instead of $v_D$, so we employ the mapping between the two to formulate the problem in terms of the mean injection velocity, which is given by:
\begin{align}
    v_i(v_D,v_{\rm th}) &= \frac{\int_{0}^\infty \frac{v}{v_{\rm th} \sqrt{  2 \pi }}\exp{\left( \frac{-\left( v-v_D \right)^2}{2 v_{\rm th}^2} \right)} \, dv}
    {\int_{0}^\infty \frac{1}{v_{\rm th} \sqrt{  2 \pi }}\exp{\left( \frac{-\left( v-v_D \right)^2}{2 v_{\rm th}^2} \right)} \, dv} \notag\\
    &=  v_D + \frac{v_{\rm th} \sqrt{\frac{2}{\pi}}\exp{\left( \frac{-v_D^2}{2v_{\rm th}^2}\right)}}{\rm{erfc}\left(-\frac{v_D}{\sqrt{2 v_{\rm th}^2}}\right)} . 
    \label{eqn:vi}
\end{align}
With this current density distribution function, there are a total of seven scalar input parameters, $q$, $m$, $D$, $\Phi_D$, $J_i$, $v_i$, and $v_{\rm th}$.
However, in the following section, the seven inputs are nondimensionalized to just three dimensionless parameters.
This problem is not known to be analytically solvable, so we utilize time-domain particle-in-cell (PIC) simulation to find steady-state solutions.

\section{Reduction to three dimensionless parameters}
\label{sec:Reduction}

In general, the input space for this 1D diode SCL current transmission problem comprises seven parameters: $q$, $m$, $D$, $\Phi_D$, $J_i$, $v_{\rm th}$, and $v_i$.
These seven dimensional parameters involve four basic units (charge, mass, length, and time), so by dimensional analysis, solutions to Eqs. {\ref{eqn:vlasovpoisson}} can be characterized by just $7-4=3$ independent dimensionless parameters.{\cite{Evans1972}} 
I.e. by choosing normalizations for the four basic units, Eqs. {\ref{eqn:vlasovpoisson}} can be nondimensionalized so that $q$, $m$, $D$, $\Phi_D$, $J_i$, $v_{\rm th}$, and $v_i$ do not appear, except in combinations of the three dimensionless parameters. 
Doing so vastly simplifies the empirical description, but maintains full generality since any desired result can be obtained from a function of just the three dimensionless parameters. 
While any reasonable normalization yields a similar simplification to a 3D parameter space, choosing normalizations that yield values near unity can help to provide physical insight and simplify empirical fitting. 
We normalize charge, mass, and length to $q$, $m$, and $D$, respectively; we normalize velocities to $v_i$, or equivalently, time to $D/v_i$. 

Unfortunately, the straightforward normalization for current density, $q v_i/D^3$, is not very meaningful. 
Instead, we normalize $J_i$ to $J_{\rm norm}$, which is carefully chosen to reduce to the cold beam SCL current in Eq. {\ref{eqn:SCLgeneralizedUnsimplified}} (i.e. $\lim_{v_{\rm th} \to 0} J_{\rm norm} = J_{cb}$ ) while being well defined for all $v_{\rm th}>0$.

To construct $J_{\rm norm}$, we first define the average injection velocity of particles that will traverse the gap in the absence of space charge ($v_{i_t}$) and the final velocity of a typical particle crossing the gap in the absence of space charge ($v_f$):

\begin{align}\
\label{eqn:vit}
    v_{i_t} &= \frac{\int_{v_r}^\infty j_i(v) \,dv}{\int_{v_r}^\infty \frac{j_i(v)}{v}  \,dv},\\
\label{eqn:vf}
    v_f &= \sqrt{v_{i_t}^2-2q\Phi_D/m}, 
\end{align}
where the reflection velocity $v_{r} = 0$ for $q\Phi_D<0$ and $v_{r} = \sqrt{ 2q\Phi_D/m}$ for $q\Phi_D>0$ (in this latter case, $v_f^2=v_{i_t}^2 - v_r^2 \geq 0$).  
Now, we use $v_f$ to define $J_{\rm norm}$:
\begin{align}\
\label{eqn:jnorm}
    J_{\rm norm} &= \frac{2m\epsilon_0}{9q}\frac{(v_i+v_f)^3}{D^2 }.
\end{align}
Comparing Eq. \ref{eqn:jnorm} to Eq. \ref{eqn:SCLgeneralizedUnsimplified} shows significant similarities. 
In particular, if $v_{i_t}$ were simply replaced with $v_i$, then these two equations would be identical.
Consequently, since $\lim_{v_{\rm th} \to 0} v_{i_t} = v_i$, $J_{\rm norm}=J_{cb}$ in the cold beam limit.
Thus $J_{\rm norm}$ has been defined so Eq. \ref{eqn:jnorm} reduces to Eq. \ref{eqn:SCLgeneralizedUnsimplified} as $v_{i_t} \to v_i$, which occurs as $v_{th} \to 0$.
With this built in, the model presented later in Sec. \ref{sec:fitting} inherits properties in the $v_{\rm th} \to 0$ limit corresponding to the cold beam solution. 

Now we construct the three dimensionless parameters: 
\begin{subequations}
\label{eqn:dimensionlessParameters}
\begin{align}
    j_* &= \frac{J_i}{J_{\rm norm}}, \label{eqn:jstar}\\
    \phi_* &= -\frac{2q}{m} \frac{\Phi_D}{v_i^2}, \label{eqn:phistar}\\
    v_* &= \frac{v_{\rm th}}{v_i}. \label{eqn:vstar}
\end{align}
\end{subequations}
Every possible scenario for the warm steady-state 1D SCL current problem with the injected current density distribution given in Eq. \ref{eqn:maxwellian} is represented by some combination of these three quantities. 

We nondimensionalize the solution output $J_t$ with $J_{\rm max}$ (the current able to cross the gap in the absence of space charge), resulting in a measure of the fractional current transmitted across the gap, $\mathcal{F}$:

\begin{align}
    J_{\rm max} &= \int_{v_{r}}^\infty j_i(v) \,dv, \label{eqn:warmBeamJ} \\
    \mathcal{F}(j_*,\phi_*,v_*) &= J_t/J_{\rm max}. \label{eqn:mathcalF}
\end{align} 
$|J_{\rm max}|$ is less than $|J_i|$ for decelerating voltages ($q\Phi_D>0$).
$\mathcal{F}\to 1$ corresponds to the weakly SCL regime (i.e. emission limited regime where $J_t \approx J_{\rm max}$). 
On the other hand, $\mathcal{F}\to 0$ corresponds to the strongly SCL regime where most of the current is reflected rather than transmitted. 
For the cold beam case ($v_*=0$), $\mathcal{F}$ is just a step function (at $j_*=1$): 
\begin{equation}
\label{eqn:coldBeamStepFunction}
    \mathcal{F}_{\rm cold}(j_*,\phi_*) = \mathcal{F}(j_*,\phi_*,0) = 
        \begin{cases}
            1 & \text{if } |J_i|<|J_{cb}| \\
            0 & \text{if } |J_i|>|J_{cb}|
        \end{cases}
\end{equation}
For a warm beam ($v_*>0$), however, the transition of $\mathcal{F}$ between 1 and 0 is gradual, reflecting the smoothness of the velocity distribution. 
Ultimately we have reduced the problem to determining the dimensionless function $\mathcal{F}(j_*,\phi_*,v_*)$ of three dimensionless parameters. 
If $\mathcal{F}(j_*,\phi_*,v_*)$ is known, then trivial scaling will yield the physical output $J_t(q, m, D, \Phi_D, J_i, v_i, v_{\rm th})$.
Since this work only addresses distributions in the form of Eq. \ref{eqn:maxwellian} with a positive drift velocity, there is a straightforward mapping between $v_i$ and $v_D$ to give the equivalent output $J_t(q, m, D, \Phi_D, J_i, v_D, v_{\rm th})$.

\section{Solution via PIC simulation for particular $(j_*,\phi_*,v_*)$}
\label{sec:PICsimulation}

We performed standard 1D electrostatic PIC simulations with the VORPAL\cite{Nieter2004} code distributed in VSim-12.\cite{Tech-X}
First, physical parameters were determined from $(j_*,\phi_*,v_*)$. 
Without loss of generality, we simulated electrons with $q=-e$ and $m=m_e$, a gap distance $D = 150\times10^{-6}$m, and drift velocity $v_D = 4.19\times10^{6}$m/s. 
Then, $v_{\rm th}$, $\Phi_D$, and $J_i$ were uniquely determined from chosen $j_*$, $\phi_*$, and $v_*$. 
Due to nondimensionalization, the particular choice of physical parameters that give the dimensionless parameters is essentially an arbitrary choice of units. 
Only cases with different dimensionless parameters are meaningfully distinct in this context.
Boundary conditions were specified by $\Phi(0,t) = 0$ and $\Phi(D,t) = \Phi_D$. 

The simulation grid cell size $\Delta x$ was chosen based on an estimation of the minimum Debye length. 
First, a conservative nominal number density was calculated as $n_{\rm nom} = 100 J_i/{q v_D}$. 
Then, $\Delta x = {\rm min}\left[\sqrt{(\epsilon_0 m v_{\rm th}^2)/(q^2)(n_{\rm nom})},D/150\right]$. 
(The upper bound of $D/150$ was determined by convergence studies of multiple cases.)
After the simulation was run, the actual minimum Debye length $\lambda_D = \sqrt{(\epsilon_0 m v_{\rm th}^2)/(q^2)n_{\rm max}}$ was calculated from the maximum steady state number density $n_{\rm max}$ to ensure the minimum Debye length was resolved.
If it wasn't, an additional simulation was run to resolve it.
For all but nine simulations used in this study, $\lambda_D / \Delta x > 2.5$.

Due to computational expense, these other nine simulations were unable to be resolved to the same degree and have $0.44 < \lambda_D / \Delta x < 1.88$. 
Fortunately, these simulation points are not clustered together in the $(j_*,\phi_*,v_*)$ parameter space, meaning the majority of their nearest neighbors are well resolved.
These few unresolved simulations yield results for $\mathcal{F}$ that differ from the linearly interpolated value from their nearest neighbors by less than 0.1. 
Therefore, these points would not significantly affect the fit, and we chose to leave them in the analysis.

All simulations were run for 50 drift velocity crossing times, i.e. $T = 50 D/v_D$.
(We observed that an apparent steady state was reached by time $4T/5$ in all considered cases.) At each timestep, a number, $n_{\rm{particles/step}}$, of equal weight macroparticles were emitted with velocities randomly sampled from the truncated Maxwellian current density distribution $j_i(v)$ in Eq. \ref{eqn:maxwellian} further truncated to four standard deviations about $v_D$, $|v-v_D|<4v_{\rm th}$. 
Note that this neglects the low velocity ($v<v_D -4v_{\rm th}$) and high velocity ($v>v_D +4v_{\rm th}$) tails of the velocity distribution.
The velocity distribution is numerically sampled at each time step and particles are injected at the plane of the boundary at a uniform, random time within $\Delta t$. 
We ensured that sufficient macroparticles were emitted per cell per step such that neglecting each tail is equivalent to omitting at most a single macroparticle representing 2\% of the transmitted current. 
There is then a maximum macroparticle velocity within the simulation, $v_{max} = [ v_{hi}^2 - \min(2q\Phi_D/m, 0)]^\frac{1}{2}$, where $v_{hi}=v_D +4v_{\rm th}$ is the largest injected velocity. 
The timestep was chosen to prevent particles with velocity $v_{max}$ from crossing a cell, i.e. $\Delta t = \Delta x/v_{max}$.  
The transmitted current density was averaged over time interval $[4T/5,T]$ to yield a measurement for $J_t$ in steady state with associated standard deviation.

For cold beams above the SCL, it is known that the transmitted current density can oscillate in the time domain. 
However, simulations have shown that introducing a small thermal velocity can heavily damp these oscillations.\cite{Lafleur2020}
If present in our simulations, any oscillations cannot have amplitudes larger than the standard deviation of $J_i$, which was on average $0.06J_i$ and does not exceed $0.11J_i$ for the measurements used in this paper. 
Differentiating oscillations from particle noise and subsequently characterizing them requires further study.
The ultimate result of the simulation is $J_t$, or equivalently, at $\mathcal{F}_{\rm sim}(j_*, \phi_*, v_*)=J_t/J_{\rm max}$.

\section{Fitting Solutions over a range of parameters}
\label{sec:fitting}

We covered the parameter space with 18 values of $\phi_*$ ($\phi_* \in $ \{-0.9, -0.7, -0.6, -0.4, -0.2, -0.15, -0.1, -0.05, 0.0, 0.05, 0.1, 0.15, 0.2, 0.3, 0.5, 0.7, 1.5, 4.0\}) and 11 values of $v_*$ ($v_* \in$ \{0.022, 0.032, 0.039, 0.045, 0.071, 0.10, 0.12, 0.14, 0.22, 0.32, 0.39\}). 
For each ($\phi_*$, $v_*$) pair we determined values of $j_*$ to cover the transition from $\mathcal{F}_{\rm sim}=1$ to $\mathcal{F}_{\rm sim}=0$ with at least nine roughly equally-spaced values of $\mathcal{F} \in [0,1]$. 
The simulation data is included in supplementary material. 
In total, 2764 simulations were used for this model.

We observed that, for small $v_*$, the time to reach steady state can increase dramatically when $j_*\approx 1$. 
Although we have not explored this in quantitative detail, we speculate that this happens because, as $v_*$ becomes small, $\mathcal{F}$ approaches a step function at $j_*=1$, where it is highly sensitive to small changes in $j_*$.  
Thus $\mathcal{F}$ is sensitive to small fractions of the injected distribution, including the particles on the verge of being reflected, which slow down to nearly zero velocity.
Reaching a self-consistent equilibrium may take many crossing times of these slowed particles. 
To avoid unconverged measurements in this near-SCL regime, simulations were not run within the window $j_* \in [0.9,1.1]$. 
This choice is supported by the observation that all simulations outside this window yield the same $\mathcal{F}_{\rm sim}$ measurement to within 0.1 with half the simulation time ($T=25D/v_D$), while test simulations within this window change $\mathcal{F}_{\rm sim}$ by more than 0.1. 
This $j_*$ window corresponding to uncertainty in $\mathcal{F}_{\rm model}$ could be shrunk by increasing the simulation run time, though further study would be needed to characterize the time evolution. 

After all simulations were run, we fit the resultant data  $\mathcal{F}_{\rm sim}(j_*, \phi_*, v_*)$ to the following model, which is essentially a sigmoid function in $j_*$, decreasing monotonically from 1 to 0 with a varying center $j_c(\phi_*, v_*)$ and two widths $j_{w1}(\phi_*, v_*)$ and $j_{w2}(\phi_*, v_*)$, each dominating in a different $j_*$ limit ($j_{w1}$ for large $j_*$ and $j_{w2}$ for small $j_*$):

\begin{widetext}
\begin{equation}
\label{eqn:warmModel}
    \mathcal{F}_{\rm model} \left(j_*, \phi_*, v_*\right) = 1 - \frac{1}{1 + \exp\left( - \frac{(j_*-j_c)}{j_{w_1}}\right) + \left(\frac{j_c}{j_*}\right)^2 \exp\left( - \frac{(j_*-j_c)}{j_{w_2}}\right)}
\end{equation}
 where $j_c(\phi_*, v_*)$, $j_{w1}(\phi_*, v_*)$, and $j_{w2}(\phi_*, v_*)$ are given by
\begin{subequations}
\label{eqn:jcandjw}
\begin{align}
    j_c(\phi_*, v_*) &= 1+ c_1 v_* \exp{ \left[ c_2(\phi_* - c_3)^3 + c_4(v_* - c_5)^2 + c_6(\phi_* - c_7) + c_8(v_* - c_9) \right] }, \label{eqn:jc}\\
    j_{w1}(\phi_*, v_*) &=  w_1 v_*^{w_8} \exp{ \left[ w_2(\phi_* - w_3)^2 + w_4(v_* - w_5)^2 + w_6(\phi_* v_* - w_7)^2  \right] }, \label{eqn:jw} \\
    j_{w2}(\phi_*, v_*) &=  w_9 v_*^{w_{16}} \exp{ \left[ w_{10}(\phi_* - w_{11})^2 + w_{12}(v_* - w_{13})^2 + w_{14}(\phi_* v_* - w_{15})^2  \right] }. \label{eqn:jw2}
\end{align}
\end{subequations}
\end{widetext}
Eq. \ref{eqn:warmModel} is defined such that $\mathcal{F}_{\rm sim}(j_c(\phi_*, v_*), \phi_*, v_*)=\frac{2}{3}$.
Values $c_{1,2,...,9}$ and $w_{1,2,...,15}$ are fitting parameters chosen to minimize the maximum pointwise difference between the model and the simulation results. 
That is, the fit minimizes the error, $\max|\mathcal{F}_{\rm sim}(j_*, \phi_*, v_*)-\mathcal{F}_{\rm model}(j_*, \phi_*, v_*)|$, over all simulated ($j_*$, $\phi_*$, $v_*$).  
Values for $c_{1,2,...,9}$ and $w_{1,2,...,16}$ are contained in Table \ref{table:fittingParams}. The error was minimized to 0.17.

We note that this model automatically incorporates limiting behavior known from the cold beam SCL in Eq. \ref{eqn:SCLgeneralized}.
Particularly, in the $v_{\rm th} \to 0$ limit, or $v_* \to 0$, we have $(j_c, j_{w1}, j_{w2}) \to (1,0,0)$. 
This yields a step function at $j_*=1$, or, equivalently, $J_i = J_{\rm norm}$. 
Due to the definition of $J_{\rm norm}$ in Eq. \ref{eqn:jnorm}, $J_{\rm norm}$ goes to  $J_{cb}$ in the cold beam limit, so Eq. \ref{eqn:warmModel} is consistent with Eq. \ref{eqn:coldBeamStepFunction}, as desired.
Additionally, $\mathcal{F}(j_* \to 0) = 1$ exactly.

Though Eq. \ref{eqn:warmModel} is monotonically decreasing in $J_i$ (and $j_*$), we note that the expected increasing monotonic behavior in $\Phi_D$ is not strictly maintained. Holding all other parameters fixed, varying $\Phi_D$ within the studied parameter space yields, at worst, a single non-monotonic dip in $\mathcal{F}_{\rm model}$ of magnitude 0.086. 
So long as accuracy greater than $\sim 0.10$ is not required, this effect can be ignored.

\begin{table*}
\caption{
\label{table:fittingParams} 
List of fitting parameters used in Eq. \ref{eqn:jcandjw}.\\
}
\begin{ruledtabular}
\begin{tabular}{ccccc}
$c_1$ & $c_2$ & $c_3$ & $c_4$ & $c_5$ \\
 1.186 & $6.028\times 10^{-3}$ & 1.233 & $-2.947\times 10^{+1}$ & $2.349\times 10^{-1}$  \\
\hline
$c_6$ & $c_7$ & $c_8$ & $c_9$ & $w_1$ \\
 $-5.461\times 10^{-4}$ & $-4.264$ & 2.736 & $3.234\times 10^{-3}$ & $1.267\times 10^{+1}$  \\
\hline
$w_2$ & $w_3$ & $w_4$ & $w_5$ & $w_6$ \\
 $-3.203\times 10^{-1}$ & $7.620\times 10^{-1}$ & $-1.399$ & $6.279\times 10^{-1}$ & $6.400\times 10^{-1}$ \\
\hline
$w_{7}$ & $w_{8}$ & $w_{9}$ & $w_{10}$ & $w_{11}$ \\
 $6.595\times 10^{-1}$ & $7.077\times 10^{-1}$ & $4.558\times 10^{+1}$ & $8.571\times 10^{-1}$ & $9.738\times 10^{-1}$  \\
\hline
$w_{12}$ & $w_{13}$ & $w_{14}$ & $w_{15}$ & $w_{16}$ \\
$-2.719\times 10^{-1}$ & $-2.523$ & $2.175\times 10^{-5}$ & $1.635\times 10^{+2}$ & 2.324 \\

\end{tabular}
\end{ruledtabular}
\end{table*}

\section{Results}
\label{sec:results}

\begin{figure*}
\includegraphics[width=0.9\textwidth]{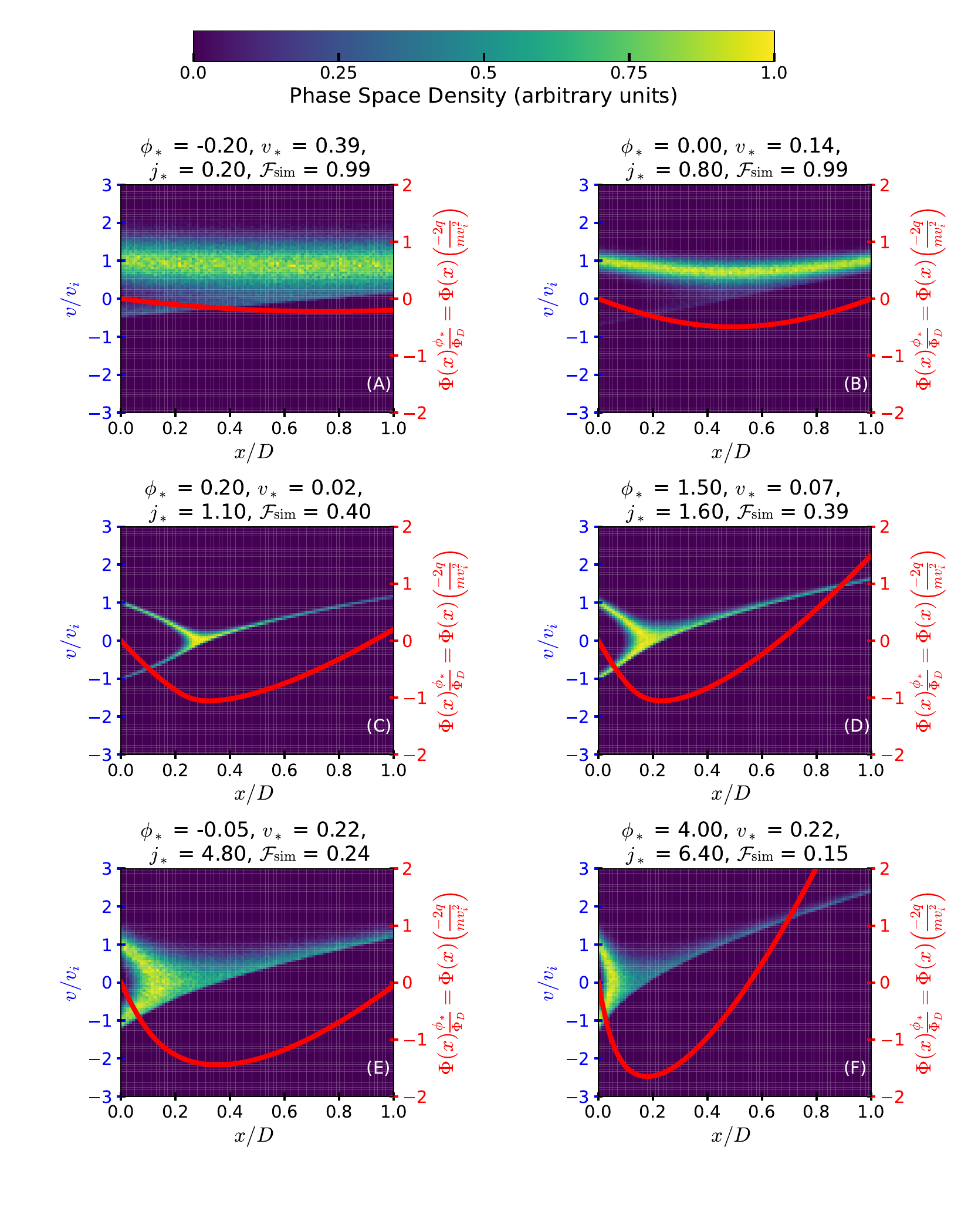}
\caption{Each subfigure (A--F) shows two quantities for a simulation with the provided dimensionless parameters: the particle phase space density (velocity and distance) from one time step at the end of the simulation (background color map) and the electric potential $\Phi(x)$ (red lines). All quantities are normalized, with distance ($x$) normalized by the gap distance ($D$), velocity normalized to the mean injection velocity ($v_i$), and the potential ($\Phi$) normalized in the same manner that $\Phi_D$ is normalized to get $\phi_*$.
}
\label{fig:intergap}
\end{figure*}

Before describing the primary result, $\mathcal{F}_{\rm sim}$, over the thousands of simulations performed, we first describe the final steady state of a few selected simulations shown in Figure {\ref{fig:intergap}}. 
Each panel shows the phase-space density in $(x,v)$-space of electrons using the blue-to-yellow color scale and the left vertical axis; overlaid in red is the potential $\Phi(x)$, using the right axis. 
These cases range from minimally space charge limited (Figure {\ref{fig:intergap}}A--B), to moderately space charge limited (Figure {\ref{fig:intergap}}C--D), to heavily space charge limited (Figure {\ref{fig:intergap}}E--F). 

In Figure {\ref{fig:intergap}}A, electrons are emitted from $x=0$ with velocities $v>0$ centered around $v_i$. The applied potential slows emitted electrons, and the curvature of $\Phi(x)$ indicates the effect of space charge. 
A small population of reflected particles are indicated by the faint green/yellow region near $x=0$ with $v<0$. 
Since $\mathcal{F}_{\rm sim}$ is not equal to 1, we know that some of these particles are reflected due to space charge. 

In Figure {\ref{fig:intergap}}B, the applied potential is zero; any space charge, no matter how small, causes $\Phi(x)$ to have a minimum below $\Phi_D =0$, and that will reflect some electrons. 
However, because $v_*$ is smaller than in {\ref{fig:intergap}}A, there are not many electrons emitted with a small enough $v$ to be reflected by the small (negative) $\Phi(x)$, and we see that a significant space charge potential forms before even $1\%$ of the current is reflected. 
Emitted electrons are visibly slowed by the potential until they are either reflected back to $x=0$, or pass the potential minimum near $x \approx 0.5 D$ and are accelerated to $x=D$.

Figures {\ref{fig:intergap}}C--F exhibit significantly more space charge effects and show the characteristic ``>''-shaped reflection pattern of slow electrons being decelerated, stopped, and then accelerated back towards the emitting plane. 
These four cases all have accelerating applied potentials.
We see that as we consider larger $j_*$, space charge increases, and the minimum $\Phi(x)$ decreases. 
This reflects more particles and lowers $\mathcal{F}_{\rm sim}$.

The model parameters that best fit $\mathcal{F}_{\rm sim}(j_*, \phi_*, v_*)$  are given in Table \ref{table:fittingParams}. 
With these parameters, Eqs. \ref{eqn:warmModel} and \ref{eqn:jcandjw} yield an approximate solution $\mathcal{F}_{\rm model}(j_*, \phi_*, v_*)$ for a range of 1D diode problems with $\phi_* \in [-0.9,4]$ and $v_* \in [0.022,0.39]$.
This corresponds to 1D SCL current problems to $0.022<v_{th} / v_i<0.39$ and $-q \Phi_D < 2m v_i^2$. 
(In the parameter space considered, $v_i$ differs from $v_D$ by at most $1\%$.)
Multiplying $\mathcal{F}_{\rm model}$ by $J_{\rm max}$ (see Sec. \ref{sec:Reduction}) yields $J_t$. 
The appendix summarizes explicitly how to find $J_t(q, m, D, \Phi_D, J_i, v_D, v_{\rm th})$ and also provides a much-simplified calculation for beam-like cases with $v_{\rm th}/v_D < 0.05$.
Python implementations of these recipes are provided in supplementary material.  

Figures \ref{fig:combinedSampleScenarios}:A-D show sample sweeps over $j_*$ for fixed ($\phi_*$, $v_*$) pairs with typical maximum errors, ranging from 0.07 to 0.09. 
Figure \ref{fig:combinedSampleScenarios}:E shows an example where $\mathcal{F}_{\rm sim}$ is fit particularly well, with a maximum error of 0.05.
Finally, we show in Figure \ref{fig:combinedSampleScenarios}:F a sweep with large error, corresponding to a maximum error of 0.16.
(The $\phi_*$ and $v_*$ choices for these panels correspond to those found in Figure {\ref{fig:intergap}} discussed below.)

We note a general trend between the relationship between $v_*$ and the derivative of $\mathcal{F}_{\rm model}$ (red curves in Figure {\ref{fig:combinedSampleScenarios}}) with respect to $j_*$; a greater $v_*$ results in a shallower slope. 
Physically, this is because a greater $v_*$ corresponds to a larger velocity spread, which means that an even greater change to the potential is required to cause the same change in $\mathcal{F}$.

\begin{figure}
\includegraphics[width=0.5\textwidth]{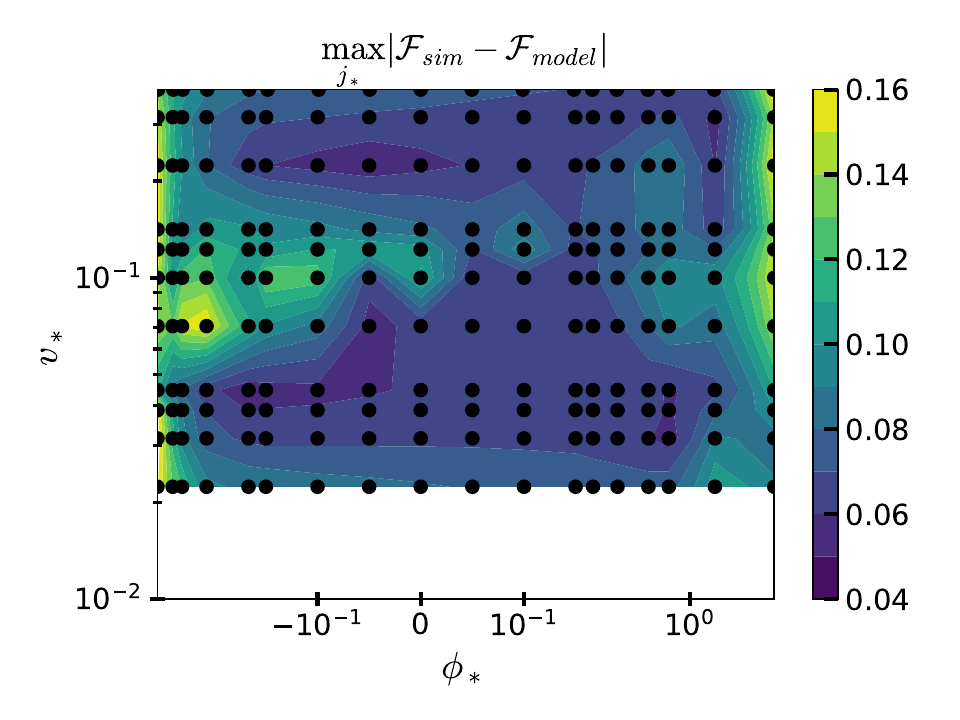}
\caption{
The maximum error in $\mathcal{F}_{\rm model}$ over the simulated $j_*$ for fixed ($v_*$, $\phi_*$). The black dots show the simulated values of ($v_*$, $\phi_*$). The worst error is $\max _{j_*}|\mathcal{F}_{sim}-\mathcal{F}_{model}| = 0.17$.
}
\label{fig:2Derror}
\end{figure}

As stated in Sec. {\ref{sec:fitting}}, the pointwise error, $|\mathcal{F}_{\rm sim}(j_*, \phi_*, v_*)-\mathcal{F}_{\rm model}(j_*, \phi_*, v_*)|$, was reduced to a maximum of 0.17 over the studied parameter space.
Figure {\ref{fig:2Derror}} shows the maximum error, $\max_{j_*}|\mathcal{F}_{\rm sim} - \mathcal{F}_{\rm model}|$, of each sweep over $j_*$ as a function of $\phi_*$ and $v_*$. This figure also shows the covered $\phi_*$ and $v_*$ in black dots.

\begin{figure*}
\includegraphics[width=0.9\textwidth]{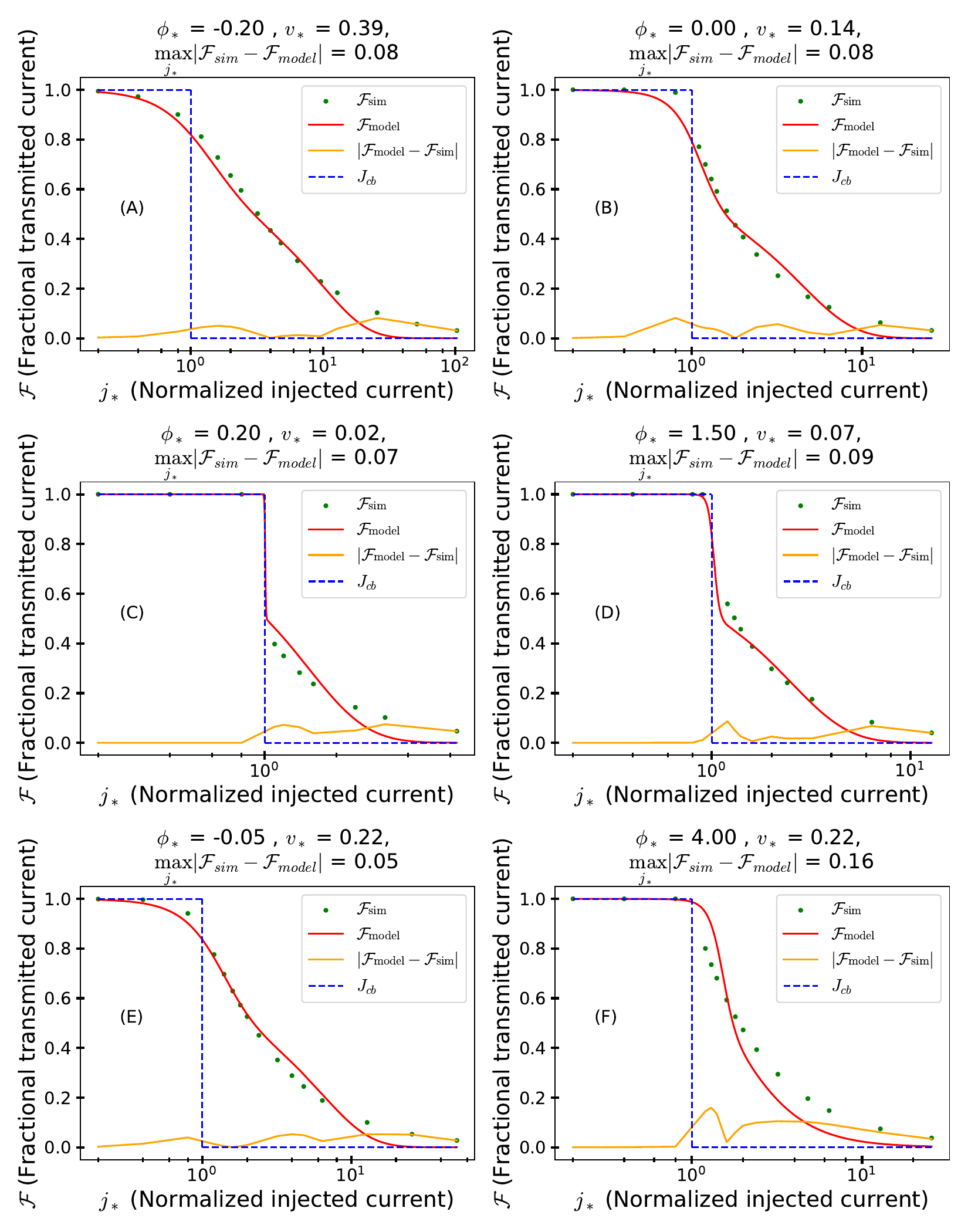}
\caption{
Each subfigure (A-F) shows values of $\mathcal{F}_{\rm sim}$ given by a simulation for fixed ($\phi_*$, $v_*$) ranging over $j_*$ (green dots). This is compared with the warm beam model $\mathcal{F}_{\rm model}$ (solid red line) and the cold beam prediction with $v_* = 0$ but the same $\phi_*$ and $v_*$ (dashed blue line exhibiting step function at $j_* = 1$). Subfigures A-D demonstrate typical error of the model. Subfigure E shows the case $j_*$ with the smallest maximum error $\max_{j_*}|\mathcal{F}_{\rm sim}(j_*, \phi_*, v_*)-\mathcal{F}_{\rm model}(j_*, \phi_*, v_*)| = 0.05$, while subfigure F shows a case with a large maximum error of $0.16$.
}
\label{fig:combinedSampleScenarios}
\end{figure*}

\section{Summary}
\label{sec:summary}

We have developed an empirical model (Eq. \ref{eqn:warmModel}) that accurately estimates the fraction, $\mathcal{F}$, of steady-state current passing through a 1D planar diode system as a function of particle and gap characteristics, with an error of less than $|\mathcal{F}{\rm sim}-\mathcal{F}{\rm model}| = 0.17$ for the transition from emission-limited regime ($\mathcal{F} \approx 1$) to SCL regime ($\mathcal{F} \approx 0$). 
This formula empirically generalizes the 1D cold beam Child-Langmuir law to the case where particles enter the diode with a finite velocity spread and applies broadly to systems within a large parameter space if the injected particle velocity distribution resembles a truncated Maxwellian (Eq. \ref{eqn:maxwellian}). 
Future research will extend the model's applicability to wider parameter ranges for $\phi_*$ and $v_*$, expand the input space to include injected Maxwellian distributions with negative drift velocities, and investigate oscillatory time-dependent behavior above the SCL.

\section*{Supplementary Material}

A Python implementation of the warm model described in this paper and detailed in the appendix is included as supplementary material. 
Additionally, the inputs and transmitted current density outputs of all 2764 simulations used in this study are also provided. 

\section*{Data Availability Statement}

The data that support the findings of this study are available from the corresponding author upon reasonable request.
\begin{acknowledgments}
This work was supported by AFOSR grant FA9550-18-1-0436.
\end{acknowledgments}

\appendix
\setcounter{secnumdepth}{0}
\section{Appendix: Recipe for Model Implementation}
\label{app:recipe}

The recipe in Table \ref{table:recipes} provides the transmitted current density in a 1D planar diode for the case of a particle distribution with a finite velocity spread, empirically extending the 1D cold beam Child-Langmuir law to a finite temperature beam.  
It includes a simplified recipe for the ``nearly'' cold beam case when the thermal velocity is finite but much smaller than the drift velocity (i.e. $v_{\rm th}>0$, but $v_{\rm th} \ll v_D$).

In the case of a nearly cold beam, simplifying approximations can be made which yield a result practically indistinguishable (result differs by less than 5\%) from the ``full'' recipe when $v_{\rm th}/v_D < 0.05$.

In both cases, we first assume a drifting Maxwellian current density distribution of the form 
$j_i(v|J_i,v_D,v_{\rm th}) =  \frac{J_i \, v}{v_{\rm norm} v_{\rm th} \sqrt{  2 \pi }} \exp{\left( \frac{-\left( v-v_D \right)^2}{2 v_{\rm th}^2} \right)}$ 
where $v_{\rm norm}$ is a normalization constant calculated in Table \ref{table:recipes}.
Other emission models described essentially by a velocity drift and spread may be approximately treated by fitting them to this functional form.
For the special case of a nearly cold beam, calculation is significantly simplified by eliminating the use of the error function, erf(x). 
The recipe in Table \ref{table:recipes} specifies how to use the model presented in this paper to calculate the transmitted steady-state current density, $J_t(q, m, D, \Phi_D, J_i, v_D, v_{\rm th})$, for both the entire covered parameter space as well as the case of a nearly cold beam.
(Note that $q$ and all current densities, $J$, are signed according to the charge carrier.)

\begin{sidewaystable*}
\vspace{3.5in}%
  \caption{Recipe for calculating the transmitted steady-state current density, $J_t(q, m, D, \Phi_D, J_i, v_D, v_{\rm th})$. 
  With fitting coefficients from Table \ref{table:fittingParams} (cf. Step 11), this model is valid for $\phi_* \in [-0.9,4.0]$ and $v_* \in [0.022,0.39]$}. 
  \label{table:recipes}

\renewcommand{\arraystretch}{2}
\begin{tabular}{|>{}p{4.75in} |>{}p{4.75in}|} 
\hline
Recipe for Model Implementation & Simplified Recipe for $v_{\rm th} \ll v_D$\\ 
\hline
1. $v_{\rm norm} = \frac{v_D}{2} \rm{erfc}\left(-\frac{v_D}{\sqrt{2}v_{\rm th}}\right) + \frac{v_{\rm th}}{\sqrt{2\pi}}\exp{\left(-\frac{v_D^2}{2v_{\rm th}^2}\right)}$ 
& 
1. Approximation: $v_{\rm norm} = v_D$
\\

2. $v_i =  v_D + v_{\rm th} \sqrt{\frac{2}{\pi}}\exp{\left( \frac{-v_D^2}{2v_{\rm th}^2}\right)} \left[\rm{erfc}\left(-\frac{v_D}{\sqrt{2 v_{\rm th}^2}}\right)\right]^{-1} $
&
2. Approximation: $v_i = v_D$
\\

3. If $q\Phi_D$ > 0, $v_r = \sqrt{2q\Phi_D/m}$, else $v_r = 0$.
&
3. If $q\Phi_D$ > 0, $v_r = \sqrt{2q\Phi_D/m}$, else $v_r = 0$.
\\

4. If $v_r = 0$, $J_{\rm max} = J_i$ , 

\qquad else $J_{\rm max} = \frac{J_i}{v_{\rm norm}} \left[ \frac{v_D}{2} \rm{erfc}\left(\frac{v_r-v_D}{\sqrt{2v_{\rm th}^2}}\right) + \frac{v_{\rm th}}{\sqrt{2\pi}}\exp{\left(-\frac{(v_r-v_D)^2}{2v_{\rm th}^2}\right)} \right]$.  
&
4. If $v_r = 0$, $J_{\rm max} = J_i$ , else $J_{\rm max} =  \frac{J_i}{2} \rm{erfc}\left(\frac{v_r-v_D}{\sqrt{2v_{\rm th}^2}}\right)$. To avoid using $\rm{erfc}$ here, 

\qquad one can use the approximation $\rm{erfc}(x) \to 1 - \rm{tanh} \left( \frac{993}{880}x + \frac{89}{880}x^3 \right) $. \cite{Vedder1987}

\\

5. $v_{i_t} = 2 v_{\rm norm}\frac{J_{\rm max}}{J_i}  \left[ \rm{erfc} \left( \frac{v_r-v_D}{\sqrt{2v_{\rm th}^2}} \right)\right]^{-1} $  
&
5. Approximation: $v_{i_t} = v_D$.
\\

6. $v_f = \sqrt{v_{i_t}^2- 2 q \Phi_D /m}$  
&
6. $v_f = \sqrt{v_{D}^2- 2 q \Phi_D /m}$
\\

\hline

\multicolumn{2}{|l|}{7. $J_{\rm norm} = \frac{2m\epsilon_0}{9 \, q}\frac{(v_i+v_f)^3}{D^2 }$  } \tabularnewline

\multicolumn{2}{|l|}{8. $j_* = \frac{J_i}{J_{\rm norm}}$ } \tabularnewline

\multicolumn{2}{|l|}{9. $\phi_* = -\frac{2q}{m} \frac{\Phi_D}{v_i^2}$ } \tabularnewline

\multicolumn{2}{|l|}{10. $v_* = \frac{v_{\rm th}}{v_i}$ } \tabularnewline

\multicolumn{2}{|l|}{11. $c_{1,2,...,9}$ and $w_{1,2,...,16}$ equal values given in Table \ref{table:fittingParams}.} \tabularnewline

\multicolumn{2}{|l|}{12. $j_c = 1+ c_1 v_* \exp{ \left( c_2(\phi_* - c_3)^3 + c_4(v_* - c_5)^2 + c_6(\phi_* - c_7) + c_8(v_* - c_9) \right) }$} \tabularnewline

\multicolumn{2}{|l|}{13. $j_{w1} =  w_1 v_*^{w_8} \exp{ \left( w_2(\phi_* - w_3)^2 + w_4(v_* - w_5)^2 + w_6(\phi_* v_* - w_7)^2  \right) }$} \tabularnewline

\multicolumn{2}{|l|}{14. $j_{w2} =  w_9 v_*^{w_{16}} \exp{ \left( w_{10}(\phi_* - w_{11})^2 + w_{12}(v_* - w_{13})^2 + w_{14}(\phi_* v_* - w_{15})^2  \right) }$} \tabularnewline

\multicolumn{2}{|l|}{15. $\mathcal{F}_{\rm model} = 1 - \frac{1}{1+\exp\left( - \frac{(j_*-j_c)}{j_{w_1}}\right) + \left(\frac{j_c}{j_*}\right)^2 \exp\left( - \frac{(j_*-j_c)}{j_{w_2}}\right)}$ } \tabularnewline

\multicolumn{2}{|l|}{16. $J_t = \mathcal{F}_{\rm model} J_{\rm max}$} \tabularnewline

\hline
\end{tabular} 
\end{sidewaystable*}


%

\end{document}